\documentclass[11pt,a4paper]{article}

\usepackage[symbol]{footmisc}

\usepackage{jheppub}

\usepackage{slashed}
\usepackage{diagbox}
\usepackage{subfigure}
\usepackage{amsmath}
\usepackage{dutchcal}
\usepackage{mathrsfs}
\usepackage{amssymb}
\usepackage{bbm}

\usepackage{amsthm}
\usepackage{amsfonts}
\usepackage{dsfont}
\usepackage{amssymb, bm}
\usepackage{array}
\usepackage{booktabs}
\usepackage{overpic}

\def\II{\hbox{{1}\kern-.25em\hbox{l}}}

\newcommand{\beq}{\begin{equation}}
\newcommand{\eeq}{\end{equation}}
\newcommand{\bal}{\begin{align}}
\newcommand{\eal}{\end{align}}

\newcommand{\nn}{\nonumber}

\newcommand{\fm}{~\mathrm{fm}}

\makeatletter
\newcommand{\newparallel}{\mathrel{\mathpalette\new@parallel\relax}}
\newcommand{\new@parallel}[2]{%
  \begingroup
  \sbox\z@{$#1T$}
  \resizebox{!}{\ht\z@}{\raisebox{\depth}{$\m@th#1/\mkern-5mu/$}}%
  \endgroup
}
\makeatother

\allowdisplaybreaks

\newcounter{MBQ}

\title{Quark Transverse Spin-Momentum Correlation of the Nucleon from Lattice QCD: The Boer-Mulders Function}

\collaboration{\noindent\makebox[\textwidth]{\bf{Lattice Parton Collaboration (LPC)}}}
\collaborationImg{\noindent\makebox[\textwidth]{\includegraphics[width=0.3\textwidth]{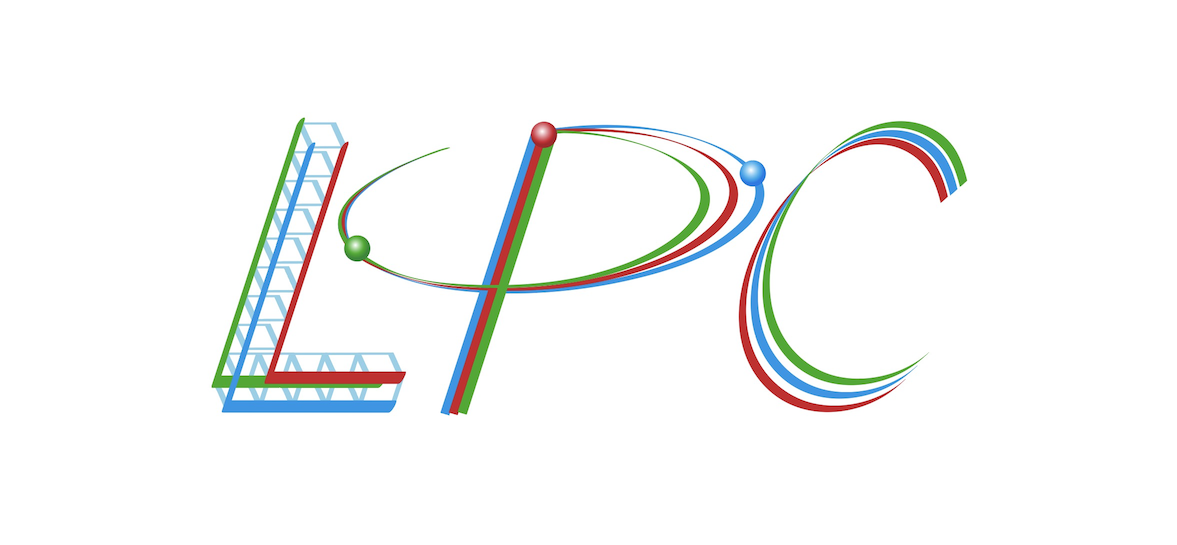}}}
\author[a,b]{Lingquan Ma}

\author[c,d]{Jun Hua}

\author[e]{Andreas Sch\"afer}

\author[f]{Hai-Tao Shu}

\author[g]{Yushan Su}

\author[h]{Peng Sun}

\author[e]{Lisa Walter}

\author[i,j]{Wei Wang}

\author[k,*]{Xiaonu Xiong
\note{Corresponding author: xnxiong@csu.edu.cn}}

\author[l,m,n,o]{Yi-Bo Yang}

\author[b,\dagger]{Jian-Hui Zhang
\note{Corresponding author: zhangjianhui@cuhk.edu.cn}}

\author[p]{Qi-An Zhang}

\affiliation[a]{Center of Advanced Quantum Studies, Department of Physics, Beijing Normal University, Beijing 100875, China}

\affiliation[b]{School of Science and Engineering, The Chinese University of Hong Kong, Shenzhen 518172, China}

\affiliation[c]{Key Laboratory of Atomic and Subatomic Structure and Quantum Control (MOE), 
Guangdong Basic Research Center of Excellence for Structure and Fundamental Interactions of Matter, 
Institute of Quantum Matter, South China Normal University, Guangzhou 510006, China}
\affiliation[d]{Guangdong-Hong Kong Joint Laboratory of Quantum Matter, 
Guangdong Provincial Key Laboratory of Nuclear Science, Southern Nuclear Science Computing Center, 
South China Normal University, Guangzhou 510006, China}

\affiliation[e]{Institut f\"ur Theoretische Physik, Universit\"at Regensburg, D-93040 Regensburg, Germany}

\affiliation[f]{Key Laboratory of Quark \& Lepton Physics (MOE) and Institute of Particle Physics, Central China Normal University, Wuhan 430079, China}

\affiliation[g]{Department of Physics, University of Maryland, College Park, MD 20742, USA}

\affiliation[h]{Institute of Modern Physics, Chinese Academy of Sciences, Lanzhou, Gansu Province 730000, China}

\affiliation[i]{INPAC, Key Laboratory for Particle Astrophysics and Cosmology (MOE), Shanghai Key Laboratory for Particle Physics and Cosmology, School of Physics and Astronomy,
Shanghai Jiao Tong University, Shanghai 200240, China}
\affiliation[j]{Southern Center for Nuclear-Science Theory (SCNT), Institute of Modern Physics, Chinese Academy of Sciences, Huizhou 516000, Guangdong Province, China}

\affiliation[k]{School of Physics, Central South University, Changsha 418003, China}

\affiliation[l]{CAS Key Laboratory of Theoretical Physics, Institute of Theoretical Physics, Chinese Academy of Sciences, Beijing 100190, China}
\affiliation[m]{School of Fundamental Physics and Mathematical Sciences, Hangzhou Institute for Advanced Study, UCAS, Hangzhou 310024, China}
\affiliation[n]{International Centre for Theoretical Physics Asia-Pacific, Beijing/Hangzhou, China}
\affiliation[o]{University of Chinese Academy of Sciences, School of Physical Sciences, Beijing 100049, China}

\affiliation[p]{School of Physics, Beihang University, Beijing 102206, China}

\abstract{We present the first lattice QCD calculation of the quark transverse spin-momentum correlation, i.e., the naive time-reversal-odd Boer-Mulders function, of the nucleon, using large-momentum effective theory (LaMET). The calculation is carried out on an ensemble with lattice spacing $a=0.098$ fm and pion mass $338$ MeV, at various proton momenta up to $2.11$ GeV. We have implemented perturbative matching up to the next-to-next-to-leading order together with a renormalization-group resummation improvement. The result exhibits a decay behavior with increasing transverse separation $b_\perp$. We also compare the results in the nucleon and pion.}


\begin{document}







\maketitle

\section{Introduction}\label{sec1}

Transverse-momentum-dependent parton distribution functions (TMDPDFs), together with generalized parton distributions (GPDs), play an essential role in mapping the three-dimensional structure of nucleons, known as the nucleon tomography program~\cite{Stefanis:2016dhq}. In contrast to GPDs which characterize the correlation between transverse position and longitudinal momentum of partons, TMDPDFs describe the correlation between transverse and longitudinal momenta of partons. 
At leading-twist accuracy, there are eight independent quark TMDPDFs which are characterized by their nucleon and quark polarizations~\cite{Boussarie:2023izj}. Among them, the naive time-reversal-odd Sivers and Boer-Mulders functions, are of special interest. While the Sivers function characterizes the correlation of quark transverse momentum and nucleon transverse polarization, the Boer-Mulders function describes the 
correlation between quark transverse polarization and momentum. 

Due to their T-odd nature, the factorization of processes involving Sivers and Boer-Mulders TMDPDFs requires the inclusion of other T-odd nonperturbative quantities such as T-odd fragmentation functions~\cite{Ji:2004xq,Bacchetta:2006tn}. This renders experimental measurements of these TMDPDFs very challenging. 
{Nevertheless, phenomenological efforts have been made to extract these functions from azimuthal angle asymmetries in semi-inclusive deep inelastic scattering (SIDIS) and Drell-Yan processes~\cite{Zeng:2024gun,Bury:2020vhj,Barone:2009hw,Lu:2009ip,Liu:2021boj}}. As a complementary tool, first-principles nonperturbative approaches like lattice QCD can also offer valuable insights into these functions. To achieve this, it is essential to calibrate them using quantities for which both state-of-the-art experiments and lattice simulations can reliably estimate precision, such as the unpolarized quark TMDPDFs. {This will enable us to trust lattice results to a similar degree, even when they cannot yet be precisely tested by experiments}. With this in mind, our long-term goal is to calculate all leading-twist TMDPDFs of hadrons using lattice QCD.

There have been exploratory lattice studies of the TMDPDFs, including the Sivers and Boer-Mulders functions, with the goal to access their moments by forming ratios of appropriate correlators~\cite{QCDSF:2007ifr,Musch:2011er,Engelhardt:2015xja,Alexandrou:2022dtc}. More recently, the development of large-momentum effective theory (LaMET)~\cite{Ji:2013dva,Ji:2014gla,Ji:2020ect} has made it possible to calculate the full TMDPDFs. Notably, a lattice calculation of the nucleon unpolarized quark TMDPDF~\cite{LatticePartonCollaborationLPC:2022myp} and the pion Boer-Mulders function~\cite{LPC:2024voq} has been carried out within the LaMET framework. In contrast to the unpolarized quark TMDPDF presented in Ref.~\cite{LatticePartonCollaborationLPC:2022myp}, a clearer decay behavior has been observed in the pion Boer-Mulders function~\cite{LPC:2024voq} with increasing transverse separations between the quark fields. This suggests that different TMDPDFs might receive rather different higher-twist contaminations, as analyzed in Ref.~\cite{Vladimirov:2020ofp}.

\begin{figure}[thbp]
\centering
\includegraphics[width=.6\textwidth]{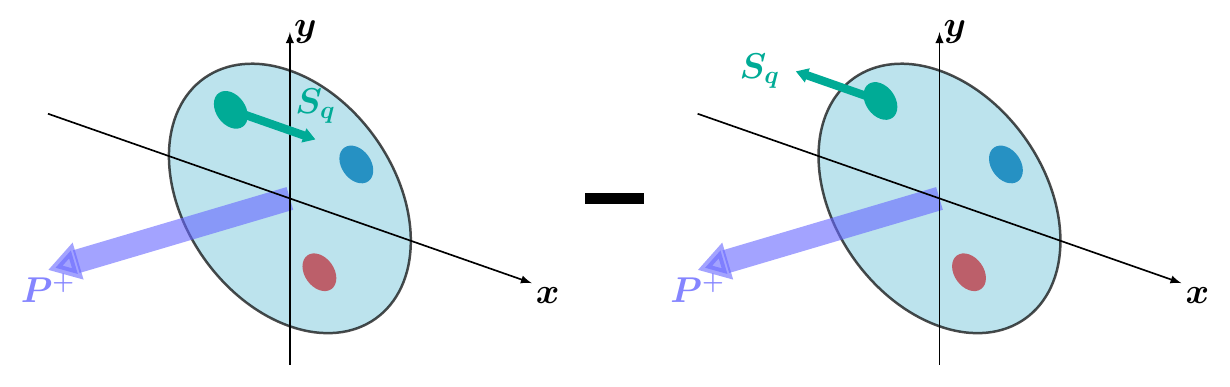}
\caption{Illustration of the physical interpretation of the Boer-Mulders TMDPDF. $S_q$ denotes the quark spin.} 
\label{fig:BM:intrp}
\end{figure}

In this work, we present a calculation of the nucleon Boer-Mulders function using LaMET. An illustration of the function is shown in Fig.~\ref{fig:BM:intrp}. The calculation is done on a lattice ensemble with lattice spacing $a=0.098$~fm and pion mass $338$ MeV. The proton momenta are up to $2.11$ GeV. The lattice matrix elements are nonperturbatively renormalized in the short-distance ratio scheme~\cite{Zhang:2022xuw} and extrapolated to large quasi-light-front (quasi-LF) distances following the same strategy as in Refs.~\cite{LatticePartonCollaborationLPC:2022myp,LPC:2024voq}. After performing a Fourier transform to longitudinal momentum space, the quasi-Boer-Mulders function is matched to the standard Boer-Mulders function using a factorization formula presented in Ref.~\cite{Ji:2019ewn}. The intrinsic soft function and Collins-Soper evolution kernel are taken from previous calculations on the same ensemble~\cite{LatticePartonLPC:2023pdv,LatticeParton:2023xdl}, and the perturbative matching kernel is implemented up to the next-to-next-to-leading order (NNLO) in~\cite{delRio:2023pse,Ji:2023pba} with a renormalization-group resummation~\cite{Su:2022fiu}. We then compare the Boer-Mulders functions in the nucleon and pion.

The rest of the paper is organized as follows: In Sec.~\ref{sec2}, we give a brief overview of the theoretical framework. In Sec.~\ref{sec3}, we present the details of our lattice calculation. In Sec.~\ref{sec4}, we discuss the numerical results of the Boer-Mulders TMDPDF in the nucleon and compare it with the same function in the pion. A summary is given in Sec.~\ref{sec5}.

\section{Theoretical Framework}\label{sec2} 
To access the quark transverse spin-momentum correlation, or the Boer-Mulders function, we start from the following subtracted quasi-TMDPDF matrix element~\cite{Ji:2014hxa,Ji:2018hvs,Ebert:2019okf,Ji:2019ewn}
\begin{align}\label{eq:MtrxElmtLatt}
    \tilde{f}(z,b_{\bot},P^z,a) &= \lim_{L\to\infty}\frac{\langle P|\hat{O}_{\sqsubset}(z,L,b_{\bot})|P\rangle}{\sqrt{Z_E(2L+z,b_\perp,a)}},
\end{align}
where $|P\rangle$ denotes an unpolarized nucleon state with momentum $P=(P^0,0,0,P^z)$, the numerator $\langle P|\hat{O}_{\sqsubset}(z,L,b_{\bot})|P\rangle$ is often referred to as the unsubtracted quasi-TMDPDF matrix element with the quark correlator
\begin{align}\label{eq:insrtd_mtrx}
    \hat{O}_{\sqsubset}(z,L,b_{\bot})&=\bar{\psi}(b_\perp \hat{n}_{\perp})i\sigma^{yt}\gamma_{5} \mathcal{W}_{\sqsubset}(z,L,b_{\bot})\psi(z\hat{n}_z),
\end{align}
where $\mathcal{W}_{\sqsubset}$ represents a staple-shaped gauge link
\begin{align}\label{eq:stplpDef}
       \notag \mathcal{W}_{\sqsubset}(b_{\bot},L,z) &= U_{z}^{\dagger}((z+L)\hat{n}_{z}+b_{\bot}\hat{n}_{\bot},b_{\bot}\hat{n}_{\bot})\\
       &\times U_{\bot}((z+L)\hat{n}_{z}+b_{\bot}\hat{n}_{\bot},(z+L)\hat{n}_{z}) U_{z}((z+L)\hat{n}_{z},z\hat{n}_{z}),\nn\\
        U_{i}(\eta+s\hat{n}_{i},\eta)&= {\cal P}{\rm exp}\Big[-ig\int_{0}^{s} dt\, \hat{n}_i\cdot A(\eta^\mu+t \hat{n}_i^\mu)\Big].
\end{align}
$\hat{n}_z, \hat{n}_\perp$ are unit vectors along the spatial $z$ and transverse directions, respectively. {The staple-shaped gauge link $\mathcal{W}_{\sqsubset}$ becomes future-pointing under large boost $P^z\to \infty$, indicating that the computed TMDPDF is relevant for SIDIS process and differs from that for Drell-Yan processes by a minus sign.} $Z_E$ is a flat rectangular Euclidean Wilson-loop along the $n_z$ direction with length $2L + z$ and width $b_\perp$,
\begin{align}\label{eq:wlp}
       Z_E(2L+z,b_\perp,a)&=\frac{1}{N_c}{\rm Tr}\langle 0|U_\perp^\dagger(-\vec L+\vec b;-b_\perp)\nonumber\times U_z^\dagger(\vec L+\vec z+\vec b_\perp;-2L-z)\nonumber\\
&\times U_{\perp}(\vec L+\vec z; b_\perp) U_z(-\vec L;2L+z)|0\rangle,
\end{align}
in which $\vec{L}=L\hat{n}_z$, $\vec{z}=z\hat{n}_z$ and $\vec{b}=b\hat{n}_\perp$. The link length $L$ is introduced to regulate the pinch-pole singularity associated with the longitudinal gauge links~\cite{Ji:2020ect}, and can be safely taken to infinity in $\tilde f$, as its dependence cancels in the ratio up to power corrections that are suppressed by $L$. 
%
In the subtracted quasi-TMDPDF, there are still residual logarithmic divergences originating from the endpoints of the quark correlator. They can be removed in the short-distance ratio scheme~\cite{Zhang:2022xuw} where the renormalization factor is defined as a subtracted quasi-TMDPDF matrix element at small $z=z_0$ and $b_\perp=b_{\perp,0}$ within perturbative region and zero momentum,
\beq
     Z_{O}(1/a,\mu) = \lim_{L\to\infty}\frac{\tilde{f}^{0}(z_0,0,b_{\bot,0},L,a)}{\sqrt{Z_{E}(2L+z_0,b_{\bot,0},a)}\;\,\tilde{h}^{\overline{\text{MS}}}_{\Gamma}(z_0,b_{\bot,0},\mu)}.
     \label{eq:Zo}
\eeq
Note that in the equation above we have included the conversion factor to $\overline{\rm MS}$ scheme $\tilde{h}_\Gamma^{\overline{\text{MS}}}$, which is expected to be the same for the quark quasi-Boer-Mulders function and for the unpolarized quark quasi-TMDPDF~\cite{Ebert:2020gxr} and takes the following form up to one-loop accuracy~\cite{Zhang:2022xuw} 
\begin{align}\label{eq:hMS}
    \tilde{h}^{\overline{\text{MS}}}_{\Gamma}(z,b_{\bot},\mu)=1+\frac{\alpha_{s}(\mu)C_{F}}{2\pi}\Big[\frac{1}{2}+\frac{3}{2}\ln\left(\frac{\mu^{2}(b_{\bot}^{2}+z^{2})e^{\gamma_{E}}}{4}\right )-2 \frac{z}{b_{\bot}}\text{arctan}\frac{z}{b_{\bot}}\Big],
\end{align}
with $\alpha_s=g^2/(4\pi)$, $C_F=4/3$. The fully renormalized quasi-TMDPDF then reads
\begin{equation}
{\tilde f}_R(z,b_{\bot},P^z,\mu)=Z_{O}^{-1}(1/a,\mu)\tilde f(z,b_\perp,P^z,a).
\end{equation}
In the renormalization factor $Z_O$, the dependence on $z_0$ and $b_{\perp,0}$ is expected to cancel. However, this cancellation is never complete due to missing higher-order perturbative contributions. To reduce the dependence on $z_0$ or $b_{\perp, 0}$, one can perform a renormalization group (RG) resummation~\cite{Su:2022fiu} from some reference scale $\mu_0$ to $\mu$, where $\mu_{0}$ is the physical scale that can be chosen as $\frac{2{e}^{-\gamma_{E}}}{\sqrt{b_\perp^{2}+z^{2}}}$. The evolved perturbative result, denoted as $\tilde{h}^{\overline{\text{MS}},\text{RGR}}_{\Gamma}(z,b_{\bot},\mu)$, is then used in Eq.~(\ref{eq:Zo}) instead of $\tilde{h}^{\overline{\text{MS}}}_{\Gamma}(z,b_{\bot},\mu)$ :
\beq\label{eq:hMSRGR}
    \tilde{h}^{\overline{\text{MS}},\text{RGR}}_{\Gamma}(z,b_{\bot},\mu) = \tilde{h}^{\overline{\text{MS}}}_{\Gamma}(z,b_{\bot},\mu_{0})\exp\Big[\int_{\alpha_{s}(\mu_{0})}^{\alpha_{s}(\mu)}\text{d}\alpha'\frac{\gamma_{F}(\alpha')}{\beta(\alpha')}\Big].
\eeq
The renormalized quasi-TMDPDF in momentum space is given by the Fourier transform
\begin{align}\label{eq:rnrml_fft}
&\tilde{f}_R(x,b_{\bot},P^z,\mu)=\int \frac{dz}{2\pi}e^{-iz(xP^{z})}\tilde f_R(z,b_{\bot},P^z,\mu).
\end{align}
%
%
%
%
It can be related to the physical Boer-Mulders TMDPDF by the following factorization formula~\cite{Ji:2019ewn}
\begin{align}\label{eq:matching}
     \tilde{f}_R(x,b_{\bot},\zeta_{z},\mu)\sqrt{S_{I}(b_{\bot},\mu)}&= \notag H_{\Gamma}\Big(\frac{\zeta_{z}}{\mu^{2}}\Big)e^{\frac{1}{2}\text{ln}\Big(\zeta_{z}/\zeta\Big)K(b_{\bot},\mu)}f(x,b_{\bot},\zeta,\mu)\nn\\
     &+\mathcal{O}\left(\frac{\Lambda_{\text{QCD}}^{2}}{\zeta_{z}},\frac{M^{2}}{P_{z}^{2}},\frac{1}{b^{2}_{\bot}\zeta_{z}}\right),
\end{align}
where $\zeta$ is the rapidity scale, and $\zeta_{z} = (2xP_{z})^{2}$. $S_I(b_\perp, \mu)$ denotes the intrinsic soft function which is associated with the
emission of soft gluons~\cite{Ji:2019ewn}, $K(b_{\bot},\mu)$ is the Collins-Soper evolution kernel~\cite{Avkhadiev:2024mgd}. Both $S_I$ and $K$ can be calculated nonperturbatively in lattice QCD \cite{LatticeParton:2020uhz,Li:2021wvl,LatticePartonLPC:2023pdv,Prokudin:2015ysa,Ebert:2018gzl,Ebert:2019tvc,Shanahan:2020zxr,Schlemmer:2021aij,Shanahan:2021tst,LatticePartonLPC:2022eev,LatticeParton:2023xdl,Avkhadiev:2023poz,Avkhadiev:2024mgd}. The $\mathcal{O}\left(\frac{\Lambda_{\text{QCD}}^{2}}{\zeta_{z}},\frac{M^{2}}{P_{z}^{2}},\frac{1}{b^{2}_{\bot}\zeta_{z}}\right)$ term denotes power corrections. $H_{\Gamma}=e^{h}$ is the hard matching kernel which has been calculated up to the next-to-next-to-leading order{~\cite{Ji:2018hvs,Ebert:2019okf,Ji:2020ect,delRio:2023pse,Ji:2023pba}}. 

One can also perform an RG resummation {to deal with the potentially large double logarithms $\sim\alpha_s^n\ln^{2n}( \zeta_z/\mu^2)$ in the hard kernel $H_\Gamma$ and it improves the reliability of the perturbative matching.} The RG resummation starts from fixed order perturbation series at the physical
scale $\zeta_z$ and evolves it to the renormalization scale $\mu$
\begin{align}
H\left(\alpha_s(\mu), \frac{\zeta_z}{\mu^2}\right)=H\left(\alpha_s\left(\sqrt{\zeta_z}\right)\right)  \exp \left\{\int_{\sqrt{\zeta_z}}^\mu \frac{d \mu^{\prime}}{\mu^{\prime}}\left[\Gamma_{\text {cusp }}\left(\alpha_s\left(\mu^{\prime}\right)\right) \ln \frac{\zeta_z}{\mu^{\prime 2}}+\gamma_C\left(\alpha_s\left(\mu^{\prime}\right)\right)\right]\right\} ,
\end{align}
where $\Gamma_{\text {cusp }}$ and $\gamma_C$ denote the cusp anomalous dimension and the single log anomalous dimension, respectively. Their explicit expressions can be found in Refs.~\cite{LatticePartonCollaborationLPC:2022myp,LPC:2024voq}. The scale $\sqrt{\zeta_z}$ is varied between $0.8\sqrt{\zeta_z}$ and $1.2\sqrt{\zeta_z}$ to estimate systematic uncertainties related to the scale choice.

\section{Lattice calculation}\label{sec3}
\subsection{Lattice setup}
In this work, we use the lattice ensemble X650 generated by the CLS collaboration~\cite{Bruno:2014jqa} {using 2+1 flavor dynamical clover fermions and tree-level Symanzik improved gauge action.} The lattice spacing is $a=0.098~\mathrm{fm}$ and pion mass is $m_\pi=338~\rm{MeV}$. Two different types of smearing are adopted to improve the signal-to-noise ratio: 
The momentum smearing source technique~\cite{Bali:2016lva} to improve the signal for a fast moving nucleon and {HYP smearing to improve the signal for non-local operators with large separations}~\cite{Hasenfratz:2001hp}.

\begin{figure}[thbp]
\centering
\includegraphics[width=0.5\textwidth]{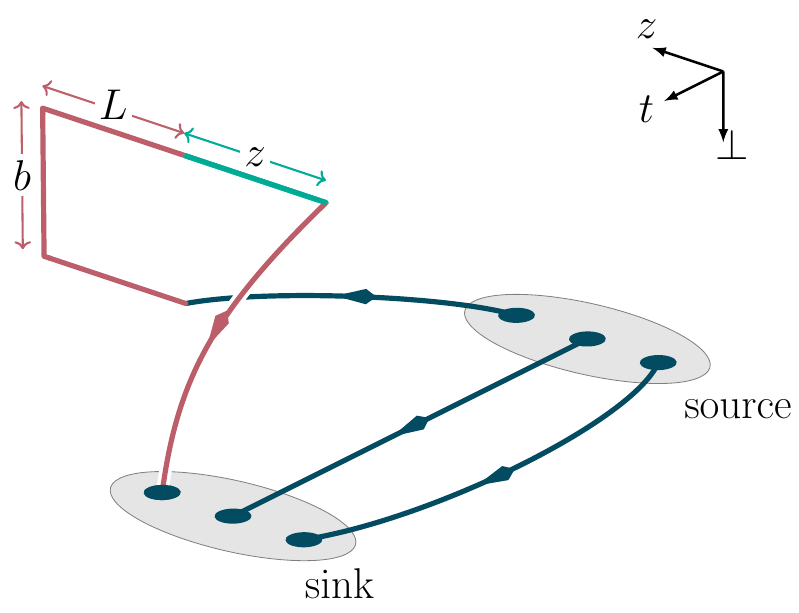}
\caption{Illustration of the nucleon three-point function. The time direction is from source to sink.} 
\label{fig:lattsetup}
\end{figure} 

We use the sequential source method with fixed sink to facilitate the
 calculation of the nucleon three-point correlator, as illustrated in Fig.~\ref{fig:lattsetup}, where we also show the staple-shaped gauge link defined in Eq.~(\ref{eq:stplpDef}) with length parameters $z,b_\perp,L$. 
 \begin{table}
\centering
\begin{tabular}{cclccccccc}
\hline
\hline
Ensemble ~&$a$(fm) ~& \ \!$L^3\times T$  ~& $m_\pi$(MeV) ~& $m_\pi L$  & $N_\mathrm{conf.}$\\
\hline
X650  ~& 0.098  ~& $48^3\times 48$  ~& 338   ~&8.1    &{{{1250}}}                                \\ 
\hline
\end{tabular}
 \caption{The lattice ensemble used in this calculation~\cite{RQCD:2022xux}}
 \label{Tab:setup_ensmbl}
\end{table}
To increase the number of measurements,
we put two sources in the temporal direction and $2, 2, 1$ sources in the $x, y, z$ directions, respectively. Details of the lattice setup and parameters are collected in Table~\ref{Tab:setup_ensmbl}. The bare matrix elements are calculated with the nucleon carrying different spatial momenta: $P^z=\{1.32, 1.58, 1.84, 2.11\}$ GeV. The bare matrix element of a nucleon at rest is also needed to extract the renormalization factor, as shown in Eq.~\eqref{eq:Zo}. In Table~\ref{tab:setup_calc}, we give details on the temporal source-sink separations, the maximum of the longitudinal separation $z_\text{max}$, transverse separation $b_\text{max}$, as well as the staple length $L$.  

\begin{table}
\centering
\begin{tabular}{ccccc}
\hline
\hline
Ensemble & $t_{\text{sep}}/a$ & $z_{\text{max}}/a$ & $b_{\perp, \text{max}}/a$ & $L/a$ \\
\hline
X650 & \{6,7,8,9,10\} & 18 & 3 & \{6,8,10\} \\
\hline
\hline
\end{tabular}
\caption{Source-sink separations $t_{\text{sep}}/a$, maximum longitudinal and transverse separations $z_{\text{max}}/a$ and $b_{\perp, \text{max}}/a$ of the quark fields in the quasi-TMDPDF defined in Eq. \eqref{eq:MtrxElmtLatt} and staple link length $L/a$ used in the analysis.}
\label{tab:setup_calc}
\end{table}

\subsection{Dispersion relation}
{To estimate the discretization effects of the X650 ensemble using nucleon's dispersion relation and extract the quasi-TMDPDF matrix element} in Eq.~\eqref{eq:MtrxElmtLatt}, we calculate the two-point function $C_{\text{2}}(P^{z},t_{\mathrm{sep}})$ defined as  
\begin{align}\label{eq:2ptdef}
    C_{\text{2}}(P^{z},t_{\mathrm{sep}})=\sum_{\vec{x}}e^{-i\Vec{P}\cdot\vec{x}}T_{u}\Big\langle{\chi(t_{\mathrm{sep}}, \Vec{x}})\Bar{\chi}(0, \vec{0})\Big\rangle,
\end{align}
where $T_{u} = (1+\gamma^{t})/2$ projects out the unpolarized nucleon state, $\chi=\epsilon^{abc}u_{a}(u_{b}^{T}C\gamma_{5}d_{c})$ is the nucleon interpolation operator. By considering the ground state and first excited state, the two-point function can be parameterized as
\begin{align}\label{eq:2ptfit}
    C_{\text{2}}(P^{z},t_{\mathrm{sep}})=c_{0}e^{-E_{0}t_{\mathrm{sep}}}(1+c_{1}e^{-\Delta Et_{\mathrm{sep}}}),
\end{align}
where $E_{0}$ denotes the ground state energy of the nucleon, $\Delta E$ is the energy shift between the first excited state and the ground state. The dispersion relation of the ground state on the lattice reads
\begin{align}\label{eq:disper}
    E_{0}(P^{z}) = \sqrt{M^{2}+c_{1}P_{z}^{2}+c_{2}P_{z}^{4}a^{2}},
\end{align}
where $c_1$ and $c_2$ take into account the discretization effect and reduce to $1$ and $0$ in the continuum limit $a\to 0$, respectively. 

\begin{figure}[thbp]
\centering
\includegraphics[width=.65\textwidth]{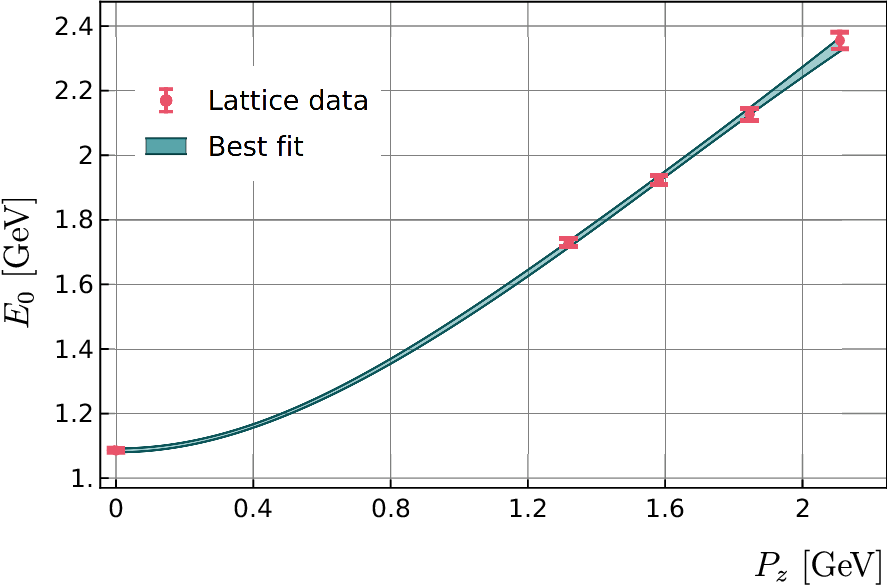}
\caption{The dispersion relation of the nucleon on the X650 ensemble.} 
\label{fig:dispersn}
\end{figure} 

By fitting two-point functions at various momenta using Eq.~\eqref{eq:2ptfit}, we obtain the nucleon's ground state energy $E_0$. $E_0$ for various $P_z$ and the fitted curve are plotted in Fig.~\ref{fig:dispersn} with $c_{1}=1.073(40)$, $c_{2}=-0.094(50)$, which are consistent with the continuum limit within $2\sigma$, {which indicates that the discretization effect is under control.} The extracted ground state nucleon mass $M$ is $1.087(6)$~GeV.

\subsection{Wilson loop from lattice calculations}\label{wlpsub}
\begin{figure}[thbp]
\centering
\includegraphics[width=.7\textwidth]{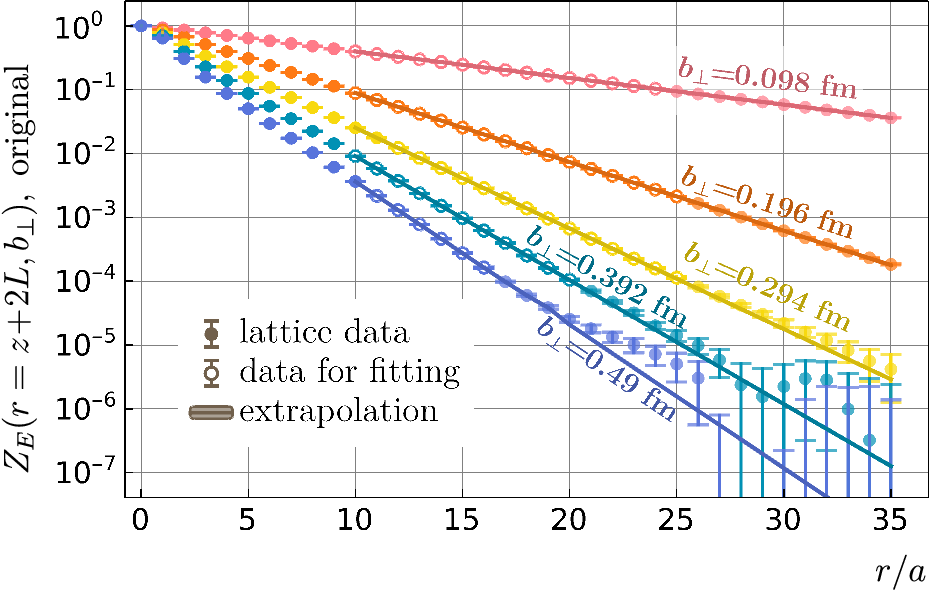}
\caption{The Wilson loop and its extrapolation with respect to $r$ and $b_{\bot}$. The points and errorbars show the original calculations of the Wilson loop, with worse signals at larger $(r,b_{\bot})$. The hollow points indicate the fit ranges for extrapolation. The bands show extrapolated results by fitting the effective energy of $Z_{E}$. We extrapolate the Wilson loop up to $r=60a$.} 
\label{fig:wilop}
\end{figure}

The Wilson loop $Z_{E}(r=2L+z,b_{\bot})$ is obtained by calculating the expectation value of a closed rectangular gauge link with side lengths $r$ and $b_{\bot}$. At large $r$ and $b_{\bot}$, the signal-to-noise ratio deteriorates, as shown in Fig.~\ref{fig:wilop}. The lattice values of $Z_{E}$ even turn negative at certain points of large $r$ and $b_{\bot}$. However, we know that the Wilson loop has to be positive. 
To avoid this, we use the same strategy as in Ref.~\cite{Zhang:2022xuw} to fit the effective energies of $Z_{E}$ which give the QCD static potential in the form $Z_E(2L+z, b_{\perp})=c(b_{\perp}) e^{-V(b_{\perp})(2L+z)}$ with $V(b_{\perp})$ being the static potential. We then extrapolate the result to large values of $r$. The extrapolated Wilson loop is shown as bands in the Fig.~\ref{fig:wilop}. In our calculation, we use the extrapolated Wilson loop to obtain the subtracted quasi-TMDPDF matrix element.

\subsection{Unsubtracted quasi-TMDPDF matrix elements from lattice calculations}

\begin{figure}[htbp]
\centering
\includegraphics[width=1\textwidth]{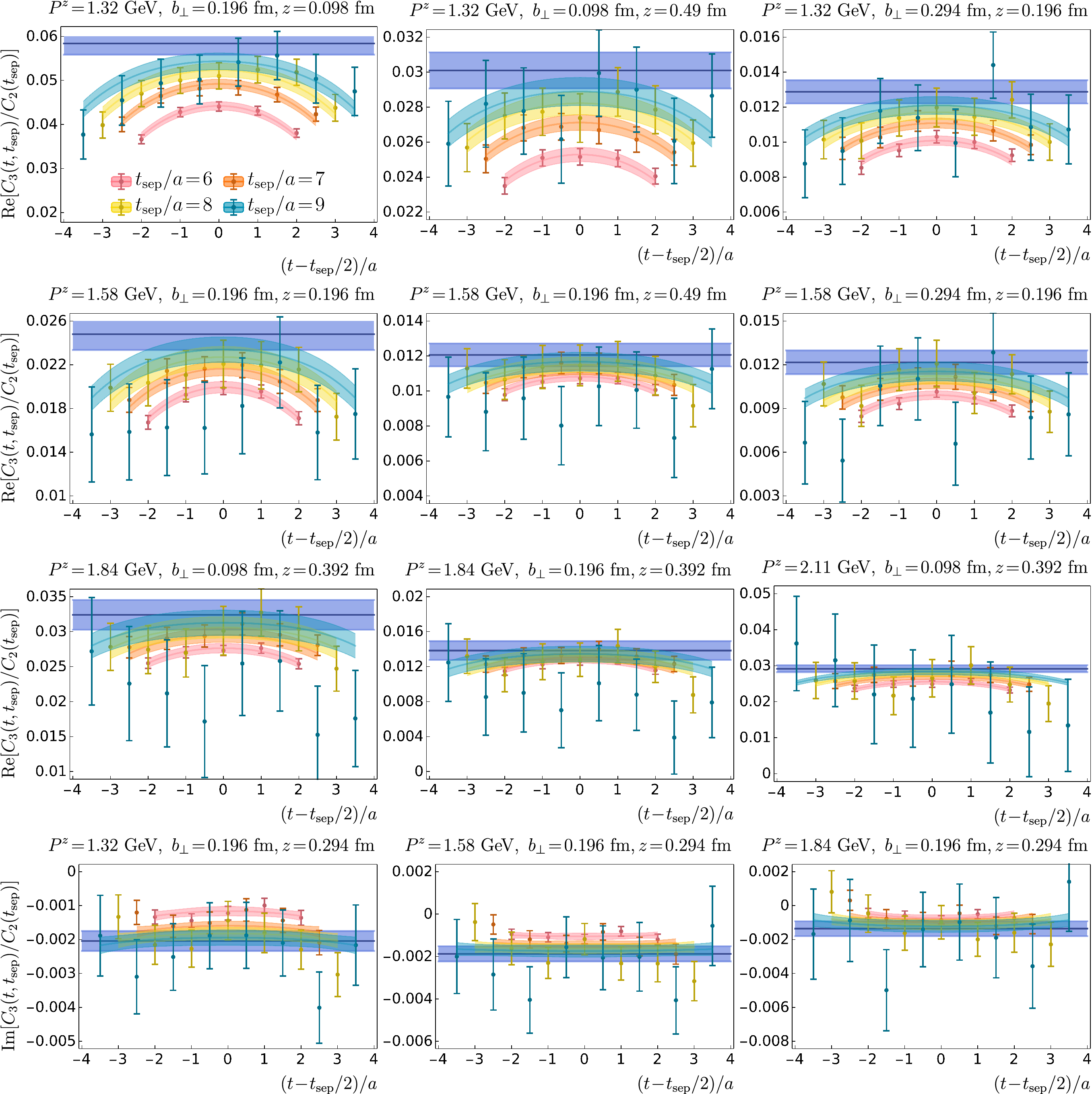}
\caption{The real and imaginary parts of the ratio $R(t,t_\text{sep})$ versus $t-t_\text{sep}/2$ for various momenta and $(z, b_{\bot})$, with $L=8a$. The straight grey band shows the central value and error of $\tilde{h}_{0}$ by fitting to the lattice data. The curved bands are $R(t,t_{\text{\text{sep}}})$ values reconstructed from the fitted parameters. {The values of $t_\mathrm{sep}$ involved in the fitting are $t_\mathrm{sep}=\{6,7,8,9\}a=\{0.588,0.686,0.784,0.882\}~\mathrm{fm}$}.} 
\label{fig:fit_c0}
\end{figure}

To extract the unsubtracted quasi-TMDPDF matrix elements, we calculate the three-piont function $C_{\text{3}}(P^{z},t_{\text{sep}},t)$ 
\begin{align}\label{eq:3ptdef}
     C_{\text{3}}(P^{z},t_{\text{sep}},t)=\sum_{\vec{x}}e^{-i\Vec{P}\cdot\vec{x}}T_{u}\Big\langle \chi(t_{\text{sep}},\Vec{x})\sum_{\vec{y}}\hat{O}(t,\vec{y})\Bar{\chi}(0,\Vec{0})\Big\rangle,
\end{align}
where $\hat{O}(t,\vec{y})$ denotes the bilinear operator $\hat{O}_{\sqsubset}(z,L,b_{\bot})$ in Eq.~\eqref{eq:MtrxElmtLatt} that is inserted at the discrete time slice $t$. {We parameterize the three-point function using a two-state fit}
\begin{align}
    C_{\text{3}}(P^{z},t_{\text{sep}},t) = c_{0}e^{-E_{0}t_{\text{sep}}}\times[\tilde{h}_{0}+c_{2}(e^{-\Delta E t}+e^{-\Delta E (t_{\text{sep}}-t)})+c_{3}e^{-\Delta E t_{\text{sep}}}],
\end{align}
{and form the ratio of the three- and two-point function}
\begin{align}\label{eq:ratio}
\frac{C_{\text{3}}}{C_{2}} = \frac{\tilde{h}_{0}+c_{2}(e^{-\Delta E t}+e^{-\Delta E (t_{\text{sep}}-t)})+c_{3}e^{-\Delta E t_{\text{sep}}}}{1+c_{1}e^{-\Delta Et_{\text{sep}}}},
\end{align}
where $\tilde{h}_{0} = \langle P|\hat{O}_{\sqsubset}(z,L,b_{\bot})|P\rangle/2E_{0}(P^{z})$ yields the desired matrix element, $E_{0}(P^{z})$ is the ground state energy. For simplicity, we denote the r.h.s. of Eq.~\eqref{eq:ratio} as $R(t,t_{\text{sep}})$ hereafter.

We use bootstrap resampling in our analysis to treat correlations in the dataset. A joint fit of the two-point function Eq.~\eqref{eq:2ptfit} and the ratio Eq.~\eqref{eq:ratio} is done after resampling data for all combinations of $(P^z,z,L,b_{\bot})$. The pionts at $t=0$ and $t=t_{\text{sep}}$ are excluded in order to reduce the contamination effects from excited states. Considering the bad signal-to-noise ratio of larger $t_{\text{sep}}$, we use $t_{\text{sep}}/a=\{6, 7, 8, 9\}$ for nonzero momenta .

As an illustrative example, we plot in Fig.~\ref{fig:fit_c0} the lattice data and fit results of the real and imaginary part of the ratio $R(t,t_{\text{sep}})$ with $L=8a$. The scattered points and error bars are lattice data corresponding to $(t_{\text{sep}},t)$ that are used in the fitting. The grey bands show the central values and errors of $\tilde{h}_{0}$. The curved bands are the reconstructed $R(t,t_{\text{sep}})$ from the fitted parameters. 

\begin{figure}[thbp]
\centering
\includegraphics[width=.5\textwidth]{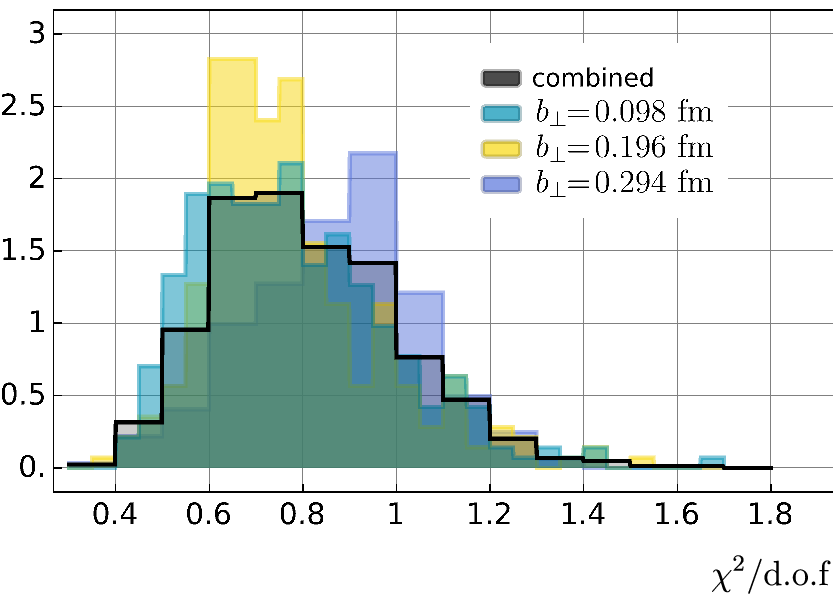}
\caption{Probability distributions of $\chi^{2}/\text{d.o.f}$. The distributions are normalized to 1.}
\label{fig:histo_chi2}
\end{figure}

To assess the quality of the combined fits, we plot histograms of $\chi^{2}/\text{d.o.f}$ of combined fits in Fig.~\ref{fig:histo_chi2}. The distributions have been normalized to 1.

\subsection{Renormalization and $L$ dependence of the subtracted quasi-TMDPDF matrix elements}\label{sub_renorm}

\begin{figure}[thbp]
\centering
\includegraphics[width=.48\textwidth]{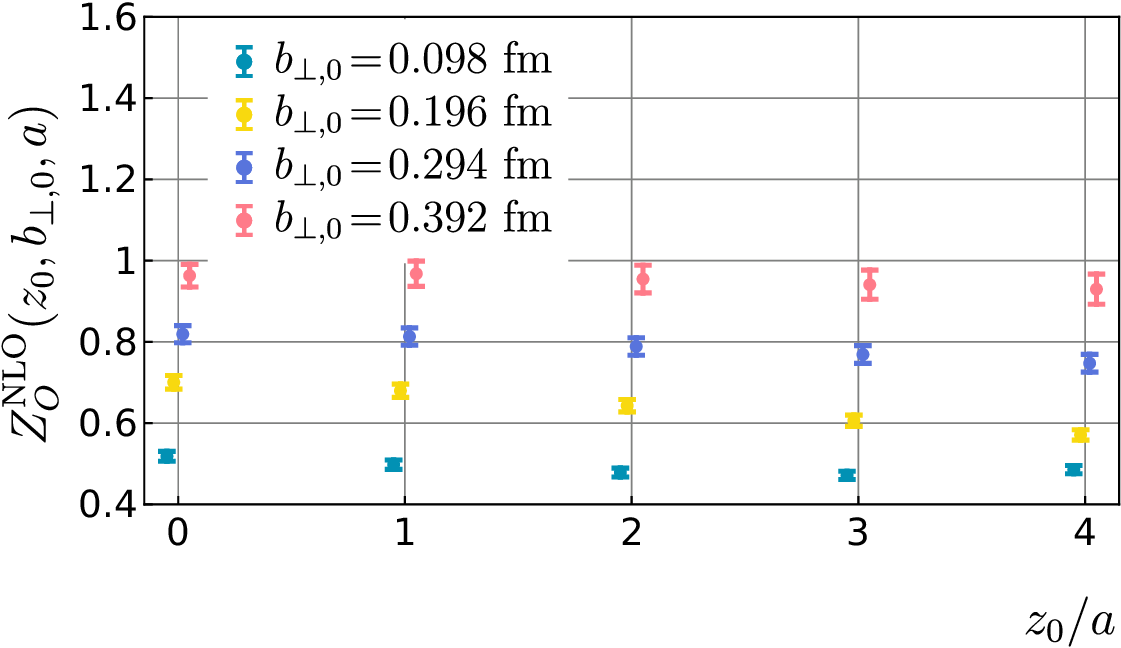}\quad
\includegraphics[width=.48\textwidth]{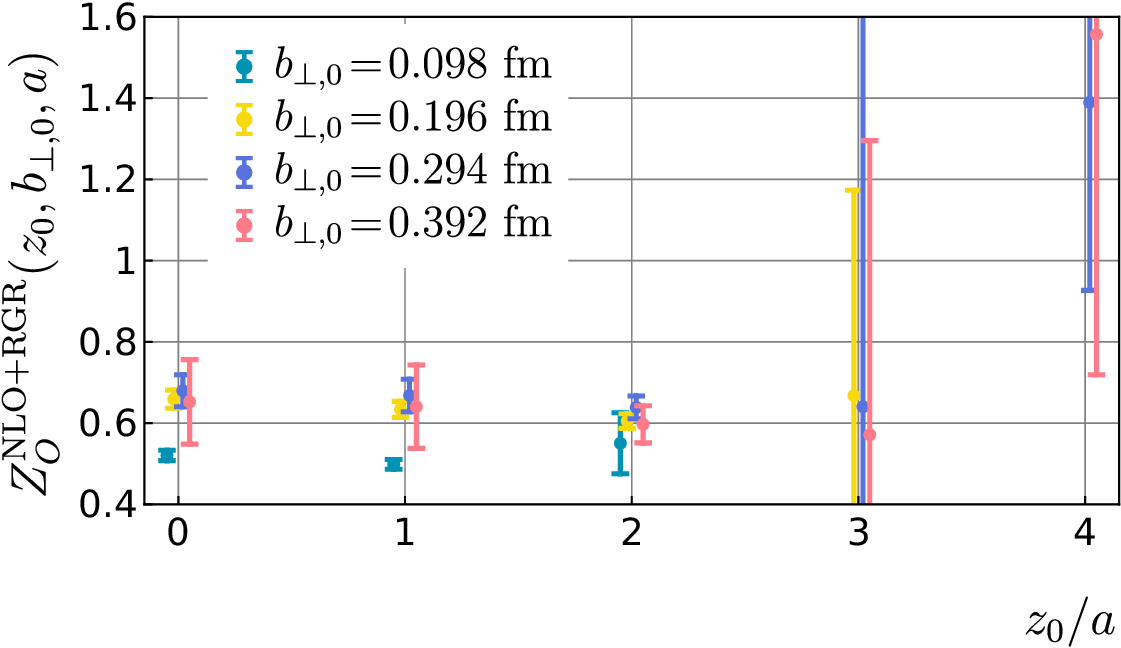}
\caption{The renormalization factor $Z_{O}$, with the NLO result as well as the NLO+RGR result at $P^z=0$ and various $b_{\bot}=b_{\perp,0}$ and $z=z_0$.}
\label{fig:nucl_nlorgr}
\end{figure}

To obtain the fully renormalized quasi-TMDPDF matrix element, we still need the logarithmic renormalization factor $Z_O$. It is computed from the zero-momentum matrix element of the same quark quasi-TMDPDF operator with $z=z_0$, $b_\perp=b_{\perp,0}$ being chosen within the perturbative region, and combined with the perturbative conversion factor. In Fig.~\ref{fig:nucl_nlorgr}, we show $Z_{O}$ obtained from the NLO perturbative result of $\tilde{h}_\Gamma^{\overline{\text{MS}}}$ in the left panel and from the NLO+RGR result in the right panel. We can see a window with $z_0\le 3 a$ and $b_{\bot,0}/a=\{2,3\}$ where the dependence of $Z_{O}$ on $z_0$ and $b_{\perp,0}$ is significantly reduced. Therefore, we use the evolved perturbative result $\tilde{h}^{\overline{\text{MS}},\text{RGR}}_{\Gamma}$ in Eq.~(\ref{eq:hMSRGR}) to calculate $Z_{O}$, where we average over $z_0$ and $b_{\bot,0}$ in the window region. The result is 0.660(16)(24), where the number in the first parenthesis is the statistical error, while the number in the second parenthesis is the systematical error estimated from varying the the RG evolution scale from $0.8\mu_{0}$ to $1.2\mu_{0}$, with $\mu_0 = 2 e^{-\gamma_E} / \sqrt{b_{\perp,0}^2 + z_0^2}$. 

\begin{figure}[thbp]
\centering
\includegraphics[width=1.\textwidth]{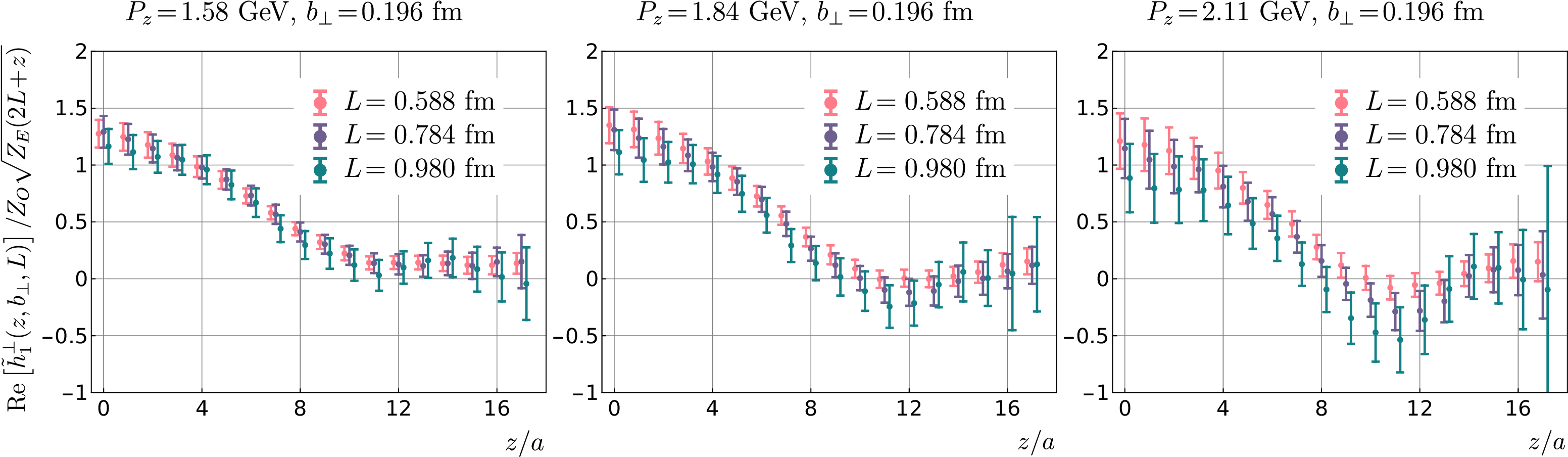}
\caption{The renormalized quasi-TMDPDF $\tilde{h}_{1}^{\bot}(z,b_{\bot},P^{z},L)$, taking $b_{\bot}=0.196\fm$ as an example. The convergence with $L$ increasing has shown in the plots, which indicates existence of an infinite $L$ limit.} 
\label{fig:subtr}
\end{figure}

After dividing the unsubtracted quasi-TMDPDF matrix element by $\sqrt {Z_{E}(2L+z,b_{\perp})}$ and $Z_O$, we find that the fully renormalized result reaches a reasonable plateau in the interval $6\leq L/a\leq 10$. 
As an example, we show Fig.~\ref{fig:subtr} the renormalized quasi-TMDPDF matrix elements for various $(P^z, b_{\bot}, L)$ versus $z$. Convergence can be observed for various $L$. The systematic error from varying $\mu_0$ has been considered in the renormalization. In the following, we take the result at $L=8a=0.784~\mathrm{fm}$ as an approximation of the result in the infinite $L$ limit.

\subsection{Large $\lambda$ extrapolation on quasi-TMDPDF in coordinate space}
\begin{figure}[thbp]
\includegraphics[width=\textwidth]{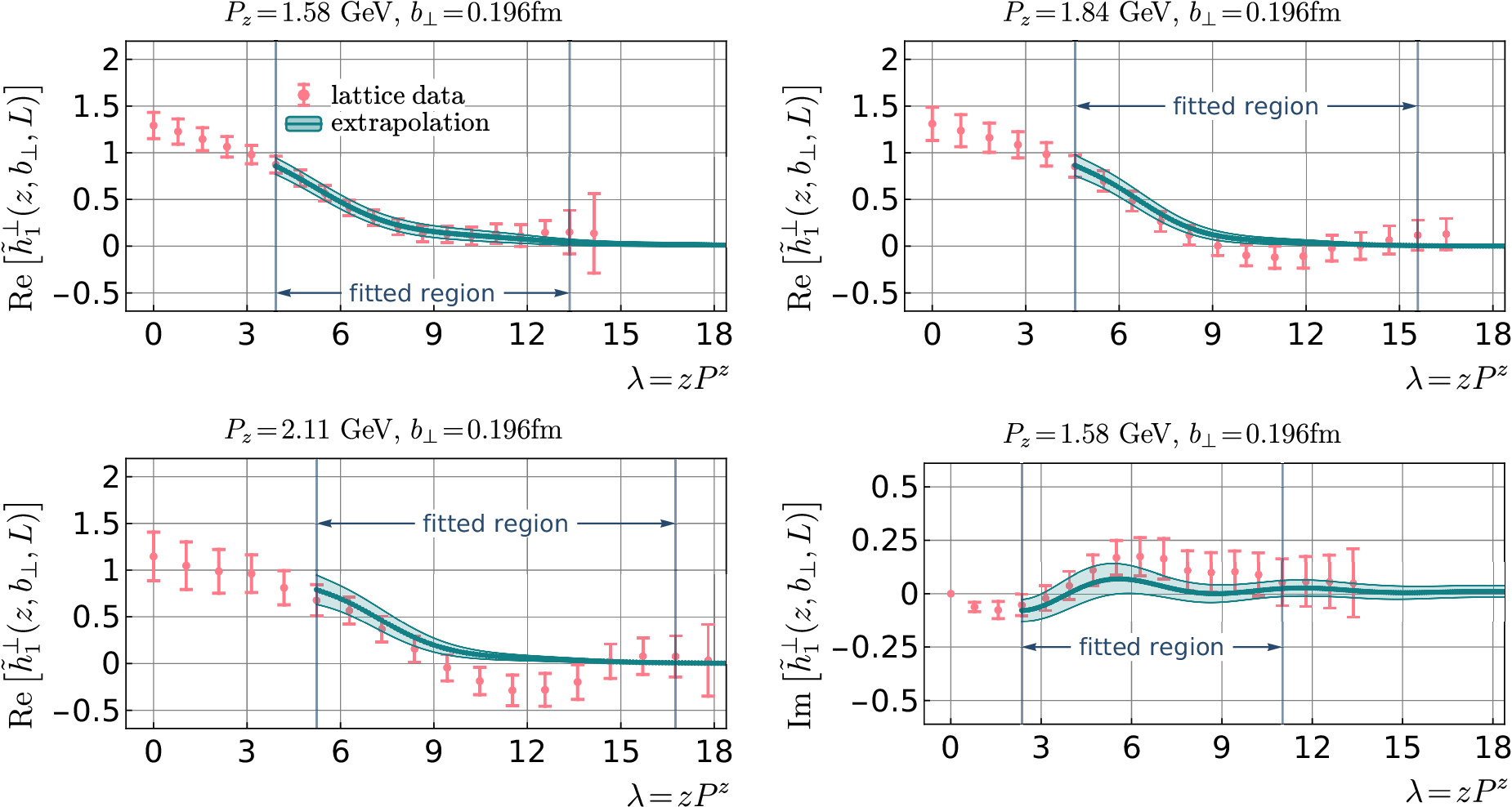}
\caption{Illustration of extrapolation for various $(P^z,b_{\bot})$. The scattered points and error bars are central values and errors of the renormalized quasi-TMDPDF $\tilde{h}_{1}^{\bot}$. The bands denote the extrapolations based on Eq.~(\ref{eq:extrpltn}). The regions used for the fitting are also indicated.}
\label{fig:extrpltn}
\end{figure}

As we can see from Fig.~\ref{fig:subtr}, the error of the quasi-TMDPDF increases rapidly at large $\lambda=zP^{z}$. In order to facilitate the subsequent Fourier transform, we perform an extrapolation to large $\lambda$ using the same ansatz as that used in Ref.~\cite{LatticePartonCollaborationLPC:2022myp}   
\begin{align}\label{eq:extrpltn}
    \tilde{h}_{\text{extra}}(\lambda) = \left[\frac{m_{1}}{(-i\lambda)^{n_{1}}}+e^{i\lambda}\frac{m_{2}}{(i\lambda)^{n_2}}\right]e^{-\lambda/\lambda_{0}},
\end{align}
where all parameters $m_{1,2}$, $n_{1,2}$ and $\lambda_{0}$ depend on $b_{\bot}$. The algebraic terms account for a power law behavior in the endpoint region of $x$, while the exponential decay is based on the expectation that the correlation length of the function (denoted as $\lambda_0$) is finite at finite momentum. 
We perform an independent extrapolation for each $b_{\bot}$. In Fig.~\ref{fig:extrpltn}, we show some examples of extrapolation for various $P^{z} = \{1.58, 1.84, 2.11\}$ GeV and $b_{\bot}=0.196\fm$. For $P^{z}=1.58$ GeV the imaginary part is also plotted. In these figures, both the original data points of $\tilde{h}^{\bot}_{1}$ and the extrapolation bands from the fitting Eq.~(\ref{eq:extrpltn}) are shown. We indicate in each plot the region chosen to fit the extrapolation form. The extrapolation results agree with lattice data in the intermediate $\lambda$ region and give smooth curves with much reduced errors for large $\lambda$ {and we replace the lattice data with $\tilde{h}_{\text{extra}}(\lambda)$ starting from the beginning of the fitted region.}

\subsection{Matching to physical Boer-Mulders TMDPDF}
With the extrapolation at large $\lambda$, we can Fourier transform the Boer-Mulders quasi-TMDPDF to momentum space and apply the matching formula in Eq.~(\ref{eq:matching}) to obtain the physical Boer-Mulders TMDPDF. 
The intrinsic soft function and Collins-Soper kernel have been calculated on X650 in~\cite{LatticePartonLPC:2023pdv}. Here we adopt these results in our matching procedure. 


\begin{figure}[thbp]
\centering
\includegraphics[width=.6\textwidth]{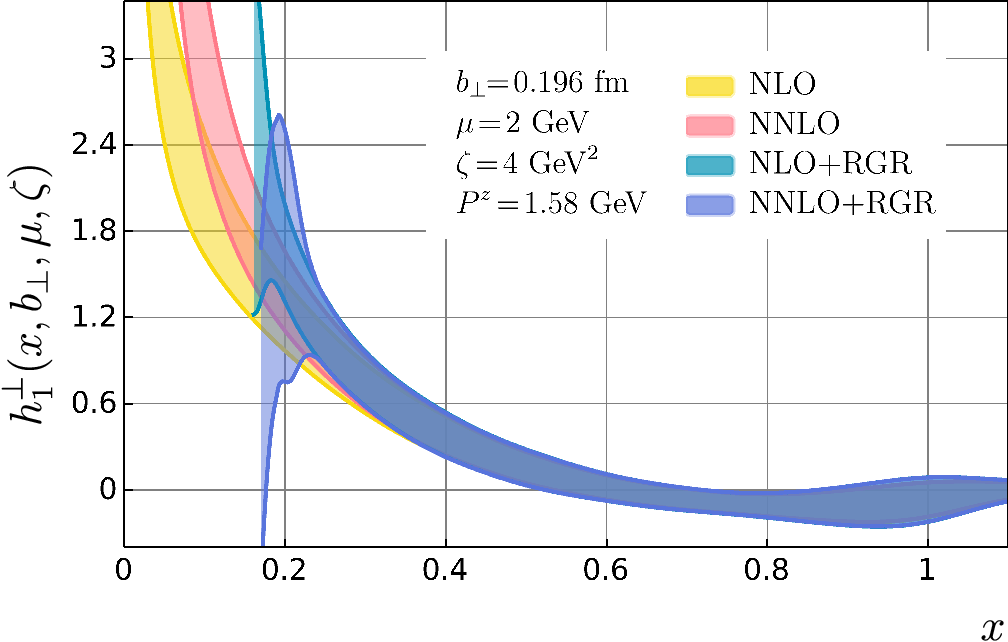}
\caption{Illustration of the Boer-Mulders function by implementing the NLO, NLO+RGR, NNLO and NNLO+RGR correction, taking $b_{\bot}=0.196\fm$, $P^z=1.58$ GeV as an example.}
\label{fig:resummtn_cmpr}
\end{figure} 

In Fig.~\ref{fig:resummtn_cmpr}, we show the effects of perturbative matching at the NLO, NLO+RGR, NNLO and NNLO+RGR, taking the $b_{\bot}=0.196\fm$, $P^z=1.58$ GeV case as an example. For the results with RGR improvement, we cut the curves at small $x$ where $2xP^{z}\lesssim\Lambda_{\mathrm{QCD}}$. As can be seen from the figure, at $x\gtrsim 0.2$ the results using the NLO, NLO+RGR, NNLO, NNLO+RGR matching are consistent with each other.

\begin{figure}[thbp]
\includegraphics[width=1\textwidth]{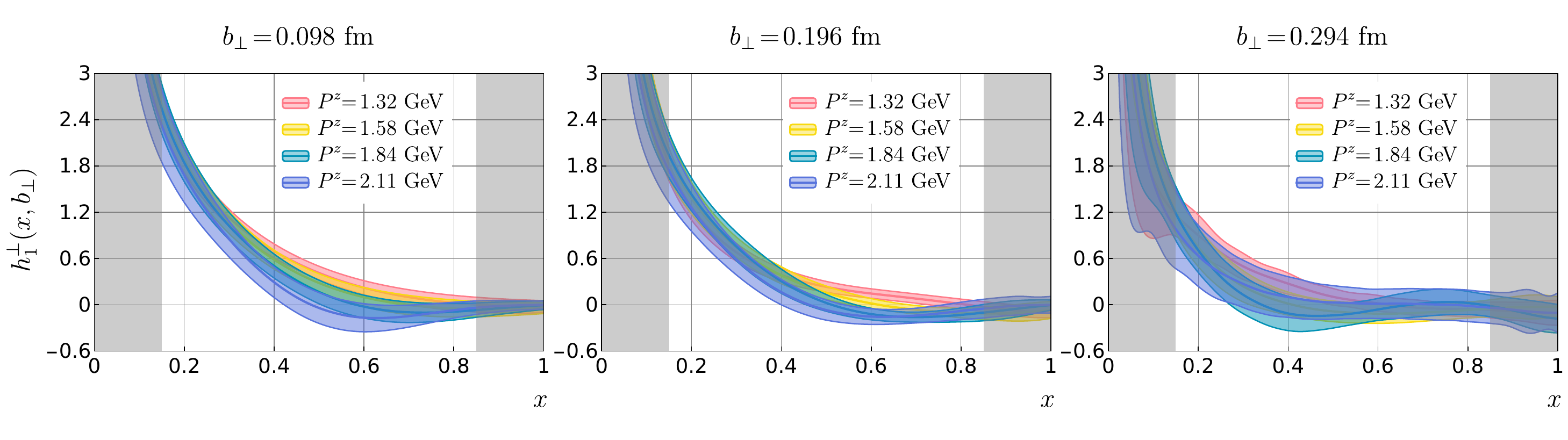}
\caption{Comparison of physical Boer-Mulders TMDPDFs at different $P^z$ and $b_{\bot}$, where the NNLO matching has been implemented. Both statistical and systematic errors have been included. Shaded regions are estimated unreliable regions where power corrections also become important.}
\label{fig:physical_cmpr_NNLO}
\end{figure}

In Fig.~\ref{fig:physical_cmpr_NNLO}, we show the momentum dependence of physical Boer-Mulders functions at different $b_\perp$ values. Given that the nucleon momenta in this calculation are not significantly larger than the nucleon mass, we do not expect a reliable extrapolation to infinite momentum. Therefore, we choose to present the results at different momenta rather than performing an infinite momentum extrapolation.
From Fig.~\ref{fig:physical_cmpr_NNLO}, we can see that the bands show a reasonable convergence behavior with increasing momentum. We only show the results up to $b_{\bot}=0.294\fm$, as it becomes difficult to control the uncertainties beyond that. The shaded regions are unreliable regions estimated from the breakdown of RGR. In these regions, power corrections in the factorization formula also become important and have to be taken into account.

\begin{figure}[thbp]
\centering
\includegraphics[width=0.65\textwidth]{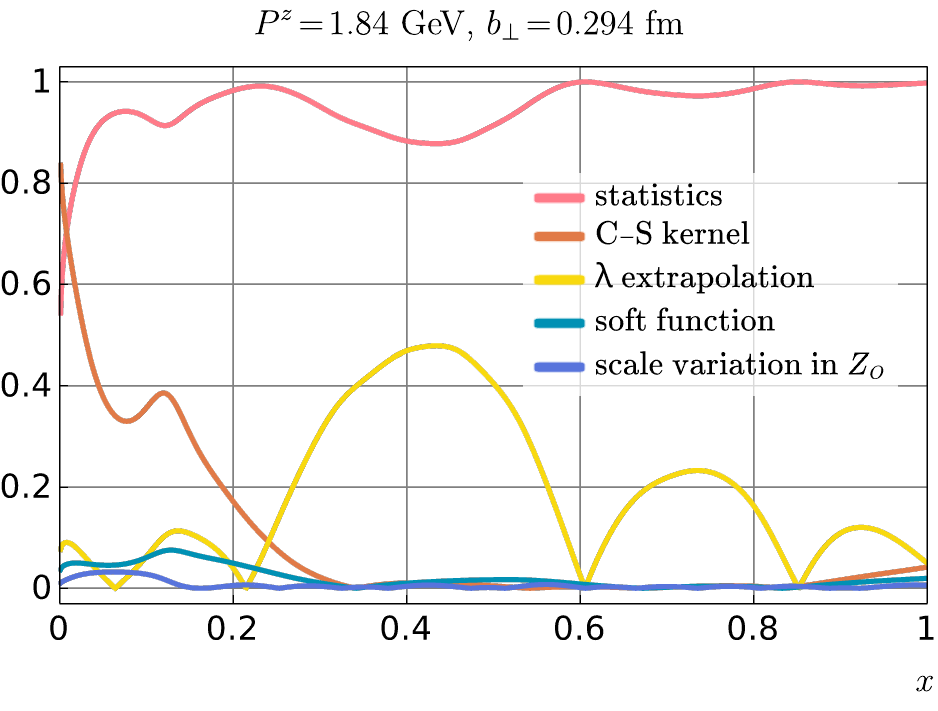}
\caption{Different sources of uncertainties, taking $P^{z}=1.84~\mathrm{GeV}$ and $b_{\bot}\!=\!0.294~\mathrm{fm}$ as an example.}
\label{fig:sys_error}
\end{figure}

The error bands in Fig.~\ref{fig:physical_cmpr_NNLO} include both statistical and systematic uncertainties, where the latter is estimated from uncertainties of the intrinsic soft function, Collins-Soper kernel, extrapolation by shifting the fitting region by $a$, and scale variation in RGR.
In Fig.~\ref{fig:sys_error}, we plot the ratio of each uncertainty and the combined uncertainty, taking the data at $P^{z}=1.84$ GeV and $b_{\bot}=0.294\fm$ as an example.

\begin{figure}[thbp]
\centering
\includegraphics[width=0.65\textwidth]{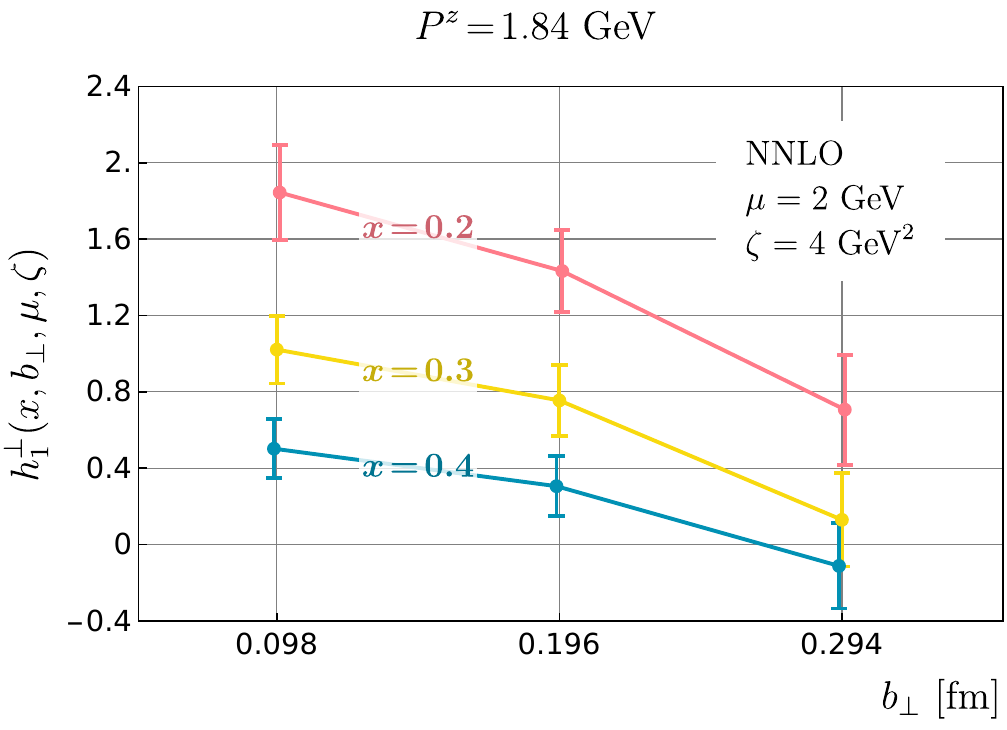}
\caption{Dependence of the nucleon Boer-Mulders function on $b_\perp$, taking $P^z = 1.84$ GeV, $x = 0.2,~0.3\text{ and }0.4$ as example.}
\label{fig:fx_vs_b}
\end{figure}

In Fig.~\ref{fig:fx_vs_b}, we show the dependence of the nucleon Boer-Mulders function on $b_\perp$. A decay trend can be observed with increasing $b_\perp$. This is in contrast with the unpolarized quark TMDPDF result in Ref.~\cite{LatticePartonCollaborationLPC:2022myp}, where the decay with $b_\perp$ is not obvious. This suggests that the Boer-Mulders function might receive smaller higher-twist contributions from the unpolarized quark TMDPDF, and keeping these contributions under control will be crucial in extracting a reliable result of the TMDPDF.


\section{Comparison with pion Boer-Mulders function}\label{sec4}


\begin{figure}[thbp]
\centering
\includegraphics[width=0.65\textwidth]{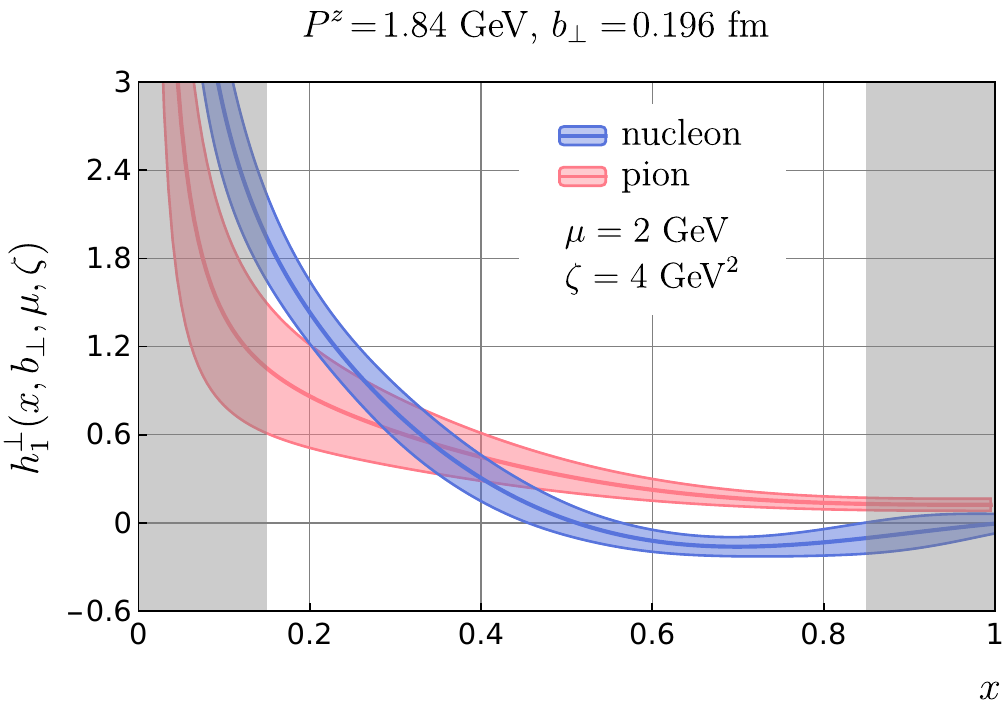}
\caption{Comparison of the Boer-Mulders functions between the nucleon and pion, taking $P^z = 1.84$ GeV, $b_\bot = 0.196\fm$ as an example. Shaded regions are estimated unreliable regions where power corrections also become important.}
\label{fig:pn_nucl_cmpr}
\end{figure}

In~Ref.\cite{LPC:2024voq}, we have calculated the Boer-Mulders TMPDF of the pion on the same ensemble X650. In Fig.~\ref{fig:pn_nucl_cmpr}, we show a comparison plot between the results of the nucleon and pion. Since no infinite momentum extrapolation has been performed in the present work due to poor signal-to-noise ratios at large nucleon momenta, our comparison is based on lattice data at the same $P^z$ and $b_\perp$. We find that the nucleon and pion Boer-Mulders functions share the same sign, as discussed in Ref.~\cite{Burkardt:2007xm}. Moreover, the function decreases more rapidly with the momentum fraction $x$ for the nucleon than for the pion. Drawing further quantitative conclusions would require a detailed study of the continuum and infinite momentum limits of the nucleon Boer-Mulders function, which we will address in future publications.

\section{Summary and Outlook}\label{sec5}
In summary, we have presented an exploratory study of the nucleon Boer-Mulders TMDPDF using LaMET. The calculation was done on an ensemble X650 with a single lattice spacing and unphysical pion mass. Our results provide preliminary insights into the Boer-Mulders function. For example, it shows a decay behavior with increasing transverse separation $b_\perp$, and shares the same sign in the nucleon and pion, with the former decaying more rapidly with the momentum fraction $x$ than the latter. More definitive conclusions would require further investigation. However, following our work, it should be relatively straightforward to identify the detailed properties of the Boer-Mulders function.

In the future, we plan to improve our calculation in the following aspects. We aim to explore the continuum, infinite momentum as well as the physical limits by collecting data on other ensembles with different lattice spacings and pion masses. To achieve this, we also need to calculate the intrinsic soft function and the Collins-Soper kernel on such ensembles. {Once this is done, we will be ready to compare our lattice results with phenomenological results obtained from fitting to experimental data}.

\begin{acknowledgments}
This work was supported by the High Performance Computing Center of Central South University. We thank the CLS Collaboration for sharing the lattices used to perform this study. The LQCD calculations were performed using the multigrid algorithm~\cite{Babich:2010qb,Osborn:2010mb} and Chroma software suite~\cite{Edwards:2004sx}. LM and JHZ are supported in part by the National Natural Science Foundation of China under grants No. 12375080, 11975051, and by CUHK-Shenzhen under grant No. UDF01002851. JH in part by the National Natural Science Foundation of China under grants No. 12205106, and by Guangdong Major Project of Basic and Applied Basic Research No. 2020B0301030008. WW is supported in part by Natural Science Foundation of China under grant No. 12125503 and  12335003. XNX is supported in part by the National Natural Science Foundation of China
under Grant No.~12275364.
YY is supported in part by NSFC grants No. 12293060, 12293062, 12435002 and 12047503, National Key R\&D Program of China No.2024YFE0109800, the Strategic Priority Research Progr of Chinese Academy of Sciences, Grant No. XDB34030303 and YSBR-101. QAZ is supported in part by the National Natural Science Foundation of China under Grant No.~12375069 and the Fundamental Research Funds for the Central Universities.
AS, HTS, PS, WW, YY and JHZ are also supported by a NSFC-DFG joint grant under grant No. 12061131006 and SCHA~458/22. PS is also supported by Strategic Priority Research Program of the Chinese Academy of Sciences under grant number XDB34030301. JH is supported in part by the National Natural Science Foundation of China under grants No. 12205106 and by Guangdong Major Project of Basic and Applied Basic Research No. 2020B0301030008.
\end{acknowledgments}

\bibliographystyle{unsrt}
\bibliography{ref}

\clearpage


\end{document}